\documentclass[useAMS,usenatbib,referee]{mn2e}
\usepackage{epsfig}

\def\ltsima{$\; \buildrel < \over \sim \;$}
\def\simlt{\lower.5ex\hbox{\ltsima}}
\def\gtsima{$\; \buildrel > \over \sim \;$}
\def\simgt{\lower.5ex\hbox{\gtsima}}

\def\h1{\ion{H}{1}\ }

\def\h2{H$_2$}
\def\coh2{CO/H$_2$}

\begin{document}

\title[A Full Loss Cone for Triaxial Galaxies]{A Full Loss Cone for Triaxial Galaxies}
\author[K. Holley-Bockelmann \& S. Sigurdsson] 
 {K.~ Holley-Bockelmann,$^{1,2}$\thanks{kellyhb@gravity.psu.edu}
 \&
 S. Sigurdsson$^{1,2}$\thanks{steinn@astro.psu.edu}\\
 $^1$ Department of Astronomy, Pennslylvania State University\\
 $^2$ Center 
for Gravitational Wave Physics, Pennsylvania State University,
University Park, PA, 16802\\}

\date{}

\pagerange{\pageref{firstpage}--\pageref{lastpage}} \pubyear{2005}

\maketitle

\label{firstpage}
 
\begin{abstract}

Stars and compact objects that plunge toward a black hole
are either
1) captured, emitting gravitational waves as the orbit decays,
2) tidally disrupted, leaving a disc of baryonic material, 
3) scattered to a large radius, where they may thereafter avoid 
encounters with the black hole or 
4) swallowed whole, contributing to black hole growth. 
These processes occur on a dynamical time, which 
implies that for a static spherically symmetric stellar system,
the loss cone is quickly emptied. However, most elliptical galaxies and 
spiral bulges are thought to be triaxial in shape. The centrophilic orbits 
comprising the backbone of a triaxial galaxy have been suggested 
as one way to keep the loss cone around a supermassive black hole 
filled with stars, stellar remnants, and intermediate mass black holes.
We investigate the evolution of the loss cone population in 
a triaxial galaxy model with high resolution N-body simulations. 
We find that enough regular orbits 
flow through angular momentum space to maintain a full loss cone 
for a Hubble time. This increases the astrophysical capture rate by several 
orders of magnitude; we derive new capture rates for the Milky Way bulge
and a triaxial M32-analogue. In the Milky Way, for example, we find that
the white dwarf capture rate can be as high as $10^{-5}$ per year, 
100 times larger
that previous estimates based on spherical models for the bulge.
This is the first step in a multiphase project that 
will explore the dynamics of supermassive black holes in realistic, 
fully self-consistent systems. The phase space content and time 
evolution of the loss cone affects supermassive black hole merger rates, 
extreme mass ratio inspiral rates, as well as the growth rate of a single
supermassive black hole due to stellar accretion, all key observables for the 
future Laser Interferometer Space Antenna (LISA).

\end{abstract}
 

\section{Introduction}

Observations suggest that supermassive black holes are a
part of nearly every galaxy center (e.g. Richstone et al. 1998), 
with masses ranging from $O(10^6)$ -
$O(10^{10}) M_\odot$. Gas accretion is thought to fuel the early 
stages of black hole growth, which may explain the tight correlation 
between black hole mass and the host galaxy's global velocity dispersion 
(Gebhardt et al. 2000a; Ferrarese \& Merritt 2000). Indeed, a supermassive black hole
may be essential to regulate its host galaxy's global structure and kinematics.

However, a single supermassive black hole (SMBH, hereafter) exacts a toll on its 
host galaxy, steepening the central cusp (Young 1980, Quinlan et al. 1995), 
and preferentially depleting the system of its low angular momentum material. 
Stars on plunging orbits interact with the SMBH and are removed from the 
galaxy by four basic processes:

\begin{enumerate}
\item Tidal Disruption 

A star is disrupted when it passes so close to the SMBH that the tidal
force exerted by the SMBH exceeds the star's own surface gravity. These 
stars are commonly thought to be the main source of the large amplitude 
X-ray outbursts seen in both active and inactive galactic nucleii
(e.g. Donley et al. 2002) as the gaseous debris settles into an accretion 
disc and falls into the  SMBH (e.g. Rees 1988, Bogdanovic et al. 2004). The tidal disruption 
radius, $r_{\rm tidal}$ depends on the stellar mass, radius, and internal 
structure:

\begin{equation}
{r_{\rm tidal}} =  {{\Bigg( {{\eta^2 {M_{\bullet} }} \over 
{{m_\star}}}\Bigg)^{1/3} } r_{\star}},
\end{equation}

where $\eta \sim 2.21$ for an incompressible, homogeneous system, and 
$\eta = 0.844$ for a main sequence star modeled with a $n=3$ polytrope 
(Magorrian \& Tremaine 1999; Sridhar \& Tremaine 1992, Diener et al. 1995). 
A G2V star would have to pass within $0.64$ AU of the Milky Way SMBH to be 
disrupted, assuming the current best Milky Way SMBH mass estimate of 
$3.7 \times 10^6 M_\odot $ (Ghez et al. 2005). 

\item Swallow Directly

For high mass SMBHs, such that $M_\bullet >\sim 10^8 M_\odot$, the tidal 
disruption radius of a main sequence star can be smaller than the 
Schwarzschild radius, $r_\bullet = 2 G M_\bullet / c^2$.
In this case, stars plunge directly into the 
SMBH, possibly without any accompanying electromagnetic event.
Since the tidal disruption radii of compact objects are much smaller, they 
can be swallowed whole by lower mass SMBHs as well: 10-30 $\%$ of the
total flux of stellar objects into a $10^6 M_\odot $ SMBH can be attributed
to compact objects and dark matter particles that have been swallowed whole 
(Zhao, Haehnelt \& Rees 2002; though see Hopman \& Alexander for a larger 
estimate).
  
\item Capture and Inspiral

Those objects that survive a close pericenter pass without 
disruption can be captured by the SMBH on a bound quasi-Keplarian orbit. 
Typically, this occurs when a white dwarf, neutron star or black hole 
(of any mass) passes 
within $few-100 M_\bullet$ of the SMBH (in $c=G=1$ units, i.e. $1.5-7.6$ AU 
for the Milky Way SMBH)\footnote{For this paper, we ignore captures from 
neutron stars}. This newly formed 
close binary has a large dynamic 
quadrupole potential and begins to emit significant gravitational radiation 
(Peters 1964) with a characteristic amplitude $h$:

\begin{equation}
{h} \sim {{4 \times 10^{-24}} {\Bigg( {{m_\star} \over {d}}\Bigg)^{1/3} } 
{\Bigg( {{M_\bullet} \over {10^6}}\Bigg)^{2/3}} {\Bigg( {{10^4} \over 
{P}}\Bigg)^{2/3}    }},
\end{equation}

where the orbital period P is measured in seconds, the stellar remnant mass 
$m_\star$ is measured in solar masses, and the distance to the source d is 
measured in Gpc. As a consequence of losing energy via gravitational 
radiation, the compact object spirals in and eventually coalesces with the 
SMBH. These extreme mass ratio inspirals (EMRIs) are a prime observational 
candidate for LISA, a planned gravitational wave telescope set to launch 
in 2017. Current estimates suggest that LISA will observe about one stellar 
mass EMRI per year within 1 Gpc of Earth (Hils and Bender 1995, 
Sigurdsson \& Rees 1997, Sigurdsson 1998, Freitag 2003, Ivanov 2002, though 
see Hopman and Alexander 2005 for a significantly  lower estimated 
rate of detection.) 

\item Ejection

Interactions between stars can induce changes in each stellar 
velocity of the order:

\begin{equation}
{\delta v_1} \sim {{4.4 \times 10^{2} {\rm {km}\over {s}}} {\Bigg( {{2 m_2} \over 
{m_1 + m_2}}\Bigg)^{{1}\over{2}} } {\Bigg( {{m_2} \over {1 M_\odot}}\Bigg)^{{1}\over{2}}} 
{\Bigg( {{1 R_\odot} \over {b}}\Bigg)^{{1}\over{2}}    }},
\end{equation}

where b is the impact parameter between the two stars, $m_1$ and $m_2$ are 
the masses of stars 1 and 2, respectively (Yu \& Tremaine 2003). It is quite 
rare to eject main sequence stars with a large enough velocity to escape the 
Milky Way entirely by this process, occurring less than once per Hubble time. 
However, if the two stars were bound as a binary, the SMBH can break 
the binary apart and eject one of the stars with $O(10^3)$ km/sec 
velocities (Hills 1988, Gould \& Quillen 2003, Yu \& Tremaine 2003). 
This process occurs more commonly 
for the Milky Way, at a rate of $10^{-5} (\eta/0.1)$ per year, where $\eta$ 
is the binary fraction (Yu \& Tremaine 2003). 

Though we have described this 3-body removal process in terms of two stars
and one SMBH, ejection is a much more efficient
process when there are two black holes and one star. If there are two 
black holes of significant mass bound at the center of the Milky Way, either 
another lower mass SMBH or an intermediate mass black hole, then the number 
of stars that can be ejected is much larger, because stars need pass only 
within the separation between the two black holes to receive energy and 
angular momentum. Given a second SMBH with a mass much smaller the the first,
Yu \& Tremaine (2003) estimate the $O(10^{-4})$ stars are ejected per year, 
with as many as 1000 stars having ejection velocities $ > 10^3$ km/sec inside
the solar radius (see also Mapelli et al. 2005).
\end{enumerate}

 Since stars are lost from the system by these 4 processes, they 
comprise the 'loss cone' of the galaxy.\footnote{ Traditionally, the term
'loss cone' has referred only to the tidal disruption process. We have 
settled on a broader definition. Tidal disruption 
is the dominant process for main sequence stars interacting with a Milky Way-scale SMBH, while 
captures are by far the dominant processes for compact objects such as white 
dwarfs, neutron stars and up to intermediate mass black holes.} The loss cone is rapidly depleted of its stellar reservoir in at most a
 few dynamical times. With a depletion timescale this rapid, it has become a
puzzle to understand how the loss cone could be anything other than empty, overall. In a static, spherical galaxy model, for example, the loss cone can only be refilled by 2-body 
relaxation, as long as 3-body scattering ejects stars permanently (e.g. Milosavljevi\'{c} \& Merritt 2003). Because the timescale for 
refilling the loss cone via 2-body relaxation can be much longer 
than a Hubble time,

\begin{equation}{
{T_{\rm relax}} \sim {2 \times 10^9 {\rm yr} \Big({{\sigma}\over {200 / {\rm km/sec}}}\Big)^3 \Big({{\rho}\over {10^6 / {\rm M_\odot/pc^3}}}\Big)^{-1}},}
\end{equation} 

the loss cone remains empty over a galaxy lifetime; this will render any type of stellar interaction with a SMBH very inefficient, at least in static, 
spherical models.

Filling a loss cone, relies on maintaining a reservoir of 
low angular momentum stars within the galaxy, so considering more 
realistic galaxy shapes may be the key. While most loss 
cone studies have concentrated on static, spherical galaxy models,
theory and observations both indicate that dark matter halos and
elliptical galaxies are at least mildly triaxial (Bak \& Statler 2000, 
Franx, Illingworth, \& de Zeeuw 1991). 
Triaxiality is present not only elliptical galaxies, but in disc galaxies 
as well: a barred galaxy is a prime example of a rotating triaxial 
ellipsoid, which has a more complex orbital structure and response to
a SMBH (Hasan \& Norman 1990, Sellwood \& Shen 2004). 
Over 70 percent of the local disc galaxy population is barred, and early 
indications suggest that this fraction may stay constant out to redshift 1 
(Jogee et al. 2004; Sheth et al. 2004). Our own galaxy hosts a bar
with a semi-major axis length of about 2 kpc seen nearly end on from 
our perspective ($\phi_{\odot -{\rm bar}} \sim 20^\circ$; see Gerhard 2001 
for a review).

Non-axisymmetry can introduce more stars to the loss cone in several ways.
First, stars in even a mildly triaxial potential move in entirely 
different orbit families than are present in a spheroid (see Figure 1). 
In particular, there are a rich variety of 
regular box and boxlet orbits that are centrophilic and comprise the 
backbone of the galaxy (Miralda-Escude \& Schwarzschild 1989). These 
centrophilic orbits can pass formally through, or very near to the 
SMBH, which make them the primary regular orbital 
component of the loss cone in a non axisymmetric galaxy.

Besides the regular box and boxlet orbits, chaotic orbits often occur 
a triaxial galaxy. Though chaotic orbits are generically in a 
triaxial galaxy, they are particularly prevalent in one that is 
embedded with a SMBH.
In fact, it is a commonly thought that SMBHs 
cannot exist in a stable triaxial galaxy, because  
black holes induce chaos in the centrophilic orbit families (Norman, May \& van Albada 1985, Gerhard \& Binney 1985, 
Miralda-Escude \& Schwarzschild 1989, Merritt \& Quinlan 1998, 
Valluri \& Merritt 1998, Holley-Bockelmann et al. 2002, Poon \& Merritt 2002). 
and drive the galaxy toward axisymmetry in a few crossing times.
Fully self-consistent, high resolution n-body simulations show, however,
that triaxiality is not entirely destroyed by a SMBH. Though SMBHs do
indeed incite chaos in box and boxlet orbits at small radii and do
rounden the region inside the radius of influence $r_{\rm inf} \equiv G M_\bullet / \sigma^2 $, most of 
the centrophilic orbits are simply scattered onto resonant (or sticky) orbits that allow 
triaxiality to persist over most of the galaxy for many Hubble times
(Holley-Bockelmann et al. 2002; see also Poon \& Merritt 2002). 

Due to their stochastic motion through phase space, chaotic orbits can
refill the loss cone as well. Stars on chaotic 
orbits encounter the SMBH once per crossing time with a pericenter 
separation that varies each time, eventually passing close enough 
to the SMBH to be captured or tidally disrupted. Merritt \& Poon (2004) 
showed that, for triaxial models where the fraction of chaotic orbits is 
substantial ( $\sim 50 \%$), the capture rate deep inside the sphere of 
influence of the SMBH meets and even exceeds that of a full loss cone for a 
spherical isotropic model. Lower energy orbits, however, experience a capture 
rate several orders of magnitude lower than a full spherical
loss cone. Despite its efficiency close to the black hole, it requires a 
chaotic orbit fraction that may not exist in real galaxies or in N-body 
generated models (Holley-Bockelmann et al. 2002).

\begin{figure}
\epsfig{file=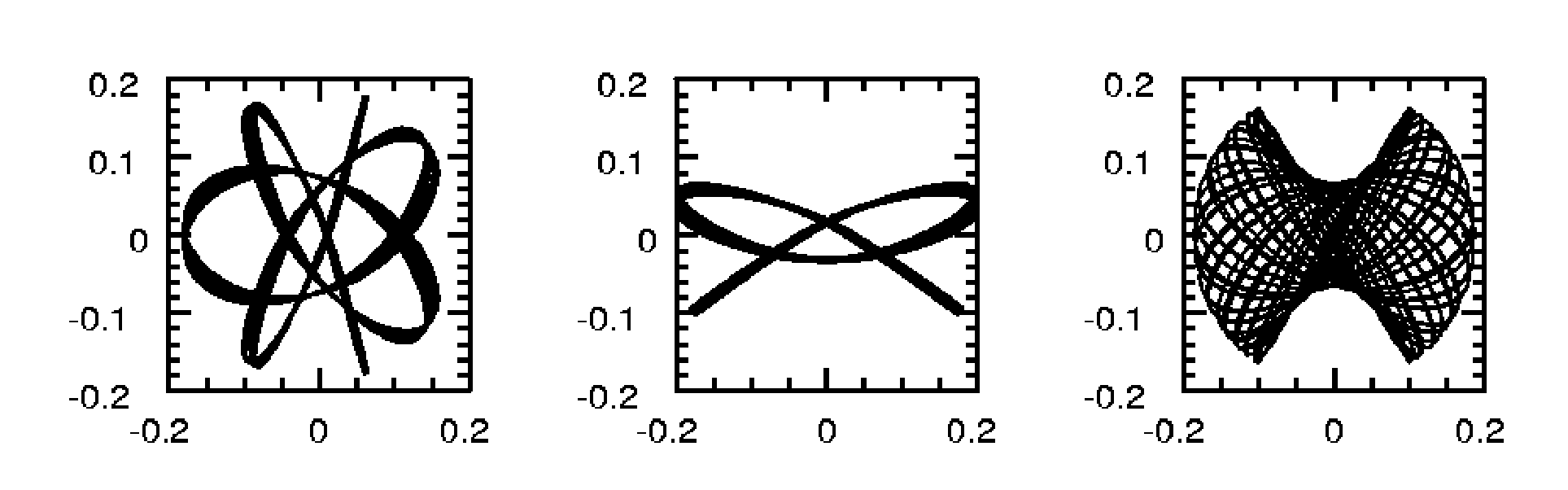, height=3in, width=7in}
\caption{Planar Centrophilic orbits in a triaxial potential with no
 SMBH. The left panel is a resonant 7:6 boxlet, the middle panel is a 
4:3 resonant boxlet (a.k.a. pretzel),  and the right panel is a non-resonant 
box. These orbits will pass through the center of the potential. 
The box orbit, in fact, has no net angular momentum and can pass formally 
through zero. In general, angular momentum is not conserved for any orbit in a triaxial potential.}
\label{fig:orbitsnobh}
\end{figure} 

In this paper we investigate a new loss cone refilling mechanism, 
uniquely present in any non-axisymmetric galaxy. Since orbits in a 
triaxial potential do not conserve angular momentum, stars can stream rapidly 
though angular momentum space. This organized flow can occur for any orbit, 
regular 
or chaotic, as a consequence of the lack of symmetry in the system which 
prevents angular momentum from being a constant of the motion. 
Since any star can flow through angular momentum space, it can be a highly 
efficient means of transporting stars through zero angular momentum. Even if
only a very small fraction of these stars are present in the loss cone, 
it may be enough to keep it filled.

Since most galaxies are so far from spherical, it is important to 
determine how more realistic galaxy models effect the structure and 
time evolution of the loss cone.
This is the first paper in a series that will investigate 
the behavior of the loss cone in general for non-spherical galaxies.
Here, we quantify how angular momentum flow refills the 
'capture and inspiral' loss cone 
in a triaxial potential and allows compact objects to be more efficiently 
captured onto orbits that emit significant gravitational radiation. We 
demonstrate this mechanism in section 3 and determine the fraction of orbits 
that can stochastically participate in 
refilling the capture and inspiral loss cone in an N-body generated triaxial 
galaxy model. We also explore the astrophysical consequences of this angular 
momentum flow on the EMRI rate, the SMBH growth rate, and the final parsec 
problem in sections 4-6. We find that this process alone can replenish the
capture and inspiral loss cone as fast as it empties, resulting in an EMRI
rate in the Milky Way that is 100 times larger than the canonical estimate for
a spherical, isotropic Milky Way bulge model.

Though we first consider angular momentum flow for all orbits in a triaxial 
potential, more traditionally diffusive processes can refill a loss cone as 
well. We include dynamical friction (the first order Fokker-Planck term)
and 'kicks' (2nd-order Fokker-Plank diffusion) using a hybrid 
SCF/Fokker-Planck code in our next paper. We are also generating a live 
$10^7$ particle triaxial model
to study the time-dependent, non-equilibrium loss cone refilling rate, 
and we are using these models to explore the decay of binary supermassive 
black holes explicitly.

\section{Numerical Techniques}

Most previous studies of the loss cone have assumed that the typical 
change in angular momenutum is small compared to the total angular momentum. 
Assuming $\Delta J << J$ allows a Fokker-Planck or perturbative approach
approach to the loss cone problem (Cohn \& Kulsrud 1978; Magorrian \& 
Tremaine 1999; Hopman \& Alexander 2005). This is useful to track 
diffusion via 2-body scattering or chaotic diffusion, 
but is not an accurate way to track the potentially large 
changes in angular momentum that may occur from this angular momentum flow.

We have decided to use n-body simulations as a testbed to
quantify the angular momentum flow. We compare two n-body generated
systems with the same final density profile and same black hole mass,
but very different shapes, one spherical and one with a varying triaxiality
from the center out, and we track the amount of angular momentum flow per orbit for each system. The procedure for generating these models is outlined
in  Holley-Bockelmann et al. 2001 and Holley-Bockelmann 
et al. 2002; we review it briefly here. The advantage of this technique is
that by using models with the same density profile and same n-body
generation technique, we can isolate the effect of angular momentum flow
has on a preexisting loss cone.

We begin with an 512,000 particle equilibrium spherical 
Hernquist model populated with a multimass scheme, so that a
particle has a mass that is roughly inversely proportional to its 
pericentric radius.
We use a parallel version of the self-consistent field (SCF) basis 
expansion code with multiple timesteps and a 4th order 
Hermite integrator to adiabatically squeeze the spherical model into 
a mildly triaxial one, then adiabatically grow a SMBH at the center 
of the potential (see Holley-Bockelmann et al. 2001; Holley-Bockelmann 
et al. 2002 for more details on the technique and model parameters). We keep
all terms in the basis expansion to $(n,l)=12,8$. This SMBH-embedded model
model has a triaxiality that varies from $a:b:c=1.0:0.95:0.92$ inside 
$r_\bullet$ to $a:b:c=1.0:0.85:0.7$ at the effective radius.

To isolate the effect of the different orbit content and integrals of 
motion in the triaxial system, we froze the potential 
of our triaxial model and followed the orbits for a Hubble time, tracking 
the change in angular momentum, pericenter distance, orbital period, eccentricity,
relaxation time and other orbital parameters. We conducted the same experiment 
using a SMBH-embedded spherical Hernquist model as a control. To decrease 
the spatial potential fluctuations, we took advantage of the 8-fold symmetry 
of our static models
to seed the final potential with mirror particles (see Holley-Bockelmann et 
al. 2001, 2002 for more details), giving the models the equivalent Poisson noise
of a 4 million particle system.






\section{Reshuffling the Loss Cone}

In this section, review some of the processes that change the orbits within the
loss cone and describe the how the loss cone reservior
can be reshuffled when orbits do not conserve angular momentum.

First, a few cautionary notes: First, when discussing the loss cone, there are 
often references to two regimes,  'diffusion' and 'pinhole'. The 'pinhole' 
regime where the loss cone width 
is small compared to $\Delta J$, and in a spherical galaxy this dominates 
at larger distances from the SMBH. Closer in, where the loss cone 
width is larger, is the 'diffusion' regime; here the
typical orbit has $\Delta J$ smaller than the loss cone
width. However, these definitions are insufficient, and even inaccurate, 
for processes that allow the orbits to change angular momentum by
large amounts. In that case, the 'pinhole' regime could dominate
all the way to the black hole for some types of orbits and not others.
For this reason, we decided not to define this angular momentum 
flow in terms of whether it acts in the pinhole regime or not. In fact,
we use this terminology sparingly. Instead, we simply quantify 
how many orbits stream through angular momentum space and how quickly they 
do so as a function of radius.

Second, in our models, we neither adjust the trajectories of 
inspiralling particles
as they begin to emit gravitational radiation nor do we extract particles from
a loss cone when it enters. It is not necessary to make this adjustment because
with only 512000 particles, the number 
of particles within the loss cone is zero. In fact, at any
given time, a galaxy with the mass of the Milky Way 
has zero occupancy within the loss cone, i.e. integrating the distribution 
function over this region of phase space indicates that expected number of 
stars within the loss cone is less than 1. The orbits which meet the 
low angular momentum criteria are both extremely rare and will interact 
with the SMBH very quickly, thus removing them from the system. 
This occurs even for a full loss cone.
So, with essentially zero occupation in the loss
cone, we follow the analytic prescriptions of Frank \& Rees (1976) and 
Sigurdsson \& Rees (1997) to define the capture and inspiral loss cone as the 
region in angular momentum space where an orbit decays
via gravitational radiation much faster than its relaxation time, 
$T_{\rm gw} < T_{\rm relax}$.


If we take the Frank and Rees (1976) approach and define the dimensionless 
angle, $\theta(r)=\sqrt{2 r_{\rm min}/3r}$,
loosely associated with the eccentricity of an orbit,
the rate at which stars are captured by two-body relaxation or large-angle scattering is:

\begin{equation}{
{R_s} = {{N_\star(r)^2 \theta_{\rm crit}^2 m_\star^2} \over 
{M_\bullet T_{\rm orb}}},}
\end{equation}

where $\theta_{\rm crit}$ for two-body relaxation is:

\begin{equation}{
{\theta_{\rm crit}} = {{\sqrt{{3} \over {2}}} {{\Big({{85 \pi } \over {24 \sqrt{2}}}\Big)}^{1/5}} {{\Big[{{M_\bullet} \over {m_\star N_\star (r)}}\Big]}^{1/5}} {{\Big({{r} \over {r_s}}\Big)}^{-1/2}}}.}
\end{equation}

And $\theta$ for large angle scattering is:

\begin{equation}{
{\theta_{\rm crit}} = {{\sqrt{{3} \over {2}}} {{\Big({{85 \pi } \over {24 \sqrt{2}}}\Big)}^{1/7}} {{\Big[{{M_\bullet} \over {m_\star N_\star (r)}}\Big]}^{1/7}} {{\Big({{r} \over {r_s}}\Big)}^{-5/{14}}}}.}
\end{equation}

The current level of discreteness noise in a $O(10^6)$ particle N-body simulation overestimates the degree of two-body relaxation and
underestimates the number of large angle scatterings in a galaxy, so it would be inaccurate to
identify the rate of depletion via these two process directly from our models.
We plan on studying these processes directly in a time-dependent
N-body simulation with better resolution,in the future. Here, we determined the width of the cone, $\theta$, 
that each of these two-body processes subtend within the phase space of
our model, and determined the rate of depletion from that area as defined in equation 5.

Even in the absence of two-body relaxation and large angle scattering, orbits in a triaxial
potential stream about in angular momentum as they obey different
integrals of motion (see figure 2). This 'reshuffles' phase space and 
systematically introduces new orbits into the loss cone. Once an orbit is 
inside the  loss cone, its apocenter shrinks rapidly, so flow {\em out} 
of the loss cone via this process is much slower than the flow in.

An orbit has completely changed its angular momentum when 
$\Delta J/J_{\rm init} = 1$. 
Figure 2 shows that, 
in total, about $0.2\%$ of the orbits in our model change 
their angular momenta by order unity in merely 1000 dynamical times, or 
$4.5 \times 10^7 {\rm years}$ at $r_{\rm inf}$ for a system scaled to the 
Milky Way. 
In figure 3, we demonstrate
how quickly each orbit reshuffles its angular momentum per dynamical time: 
$\Delta J/J_{\rm init}\big|_{t_{\rm dyn}}$ for 4 regions of total energy 
in the model.

For each particle in our simulation, we calculate the time it would take to 
change angular momentum by order unity, $t_{\rm reshuffle}$. We determine this 
 directly from the change in angular momentum per dynamical time for
each orbit within the frozen triaxial potential. By keeping track of 
the how rapidly each orbits flows through phase space, we can estimate the rate any 
particular loss cone is refilled due to this process alone:

\begin{equation}{
{R_{\rm reshuffle}} = {{{{N(r)} \over {t_{\rm reshuffle}}}}},}
\end{equation}

where $t_{\rm reshuffle}$ is:

\begin{equation}{
{t_{\rm reshuffle}} = { {{J_{\rm init} \over {\Delta J}}{\bigg| _{t_{\rm dyn}}} t_{\rm dyn} {\Bigg[ {1 - \ \  {\theta_{\rm crit}^2}}\Bigg]}}}.}
\end{equation}

Note that if a model had equal-mass particles, N(r) would be a sum over the 
number of particles found in a radial bin. However, recall that we generated
 our 
numerical model with a 'multimass' technique (see Sigurdsson et al 1998; 
Holley-Bockelmann et al. 2001), assigning more particles with smaller 
individual masses to the center. These different masses change the weight of each individual 
particle in the rate estimate, making $N(r)= \sum_{i=1}^{n} m(i)/M(r)$, where $M(r)$ is the
mass of the model at radius $r$ and $m(i)$ is the particle mass. Since this 
equation simply records how long a particular orbit takes to completely scramble its angular momentum, it holds for any density profile, any galaxy shape, and
any type of orbit. In fact, in the trivial case of one star moving in a closed orbit about a point mass, $t_{\rm reshuffle}$ is infinite.

\begin{figure}
\epsfig{file=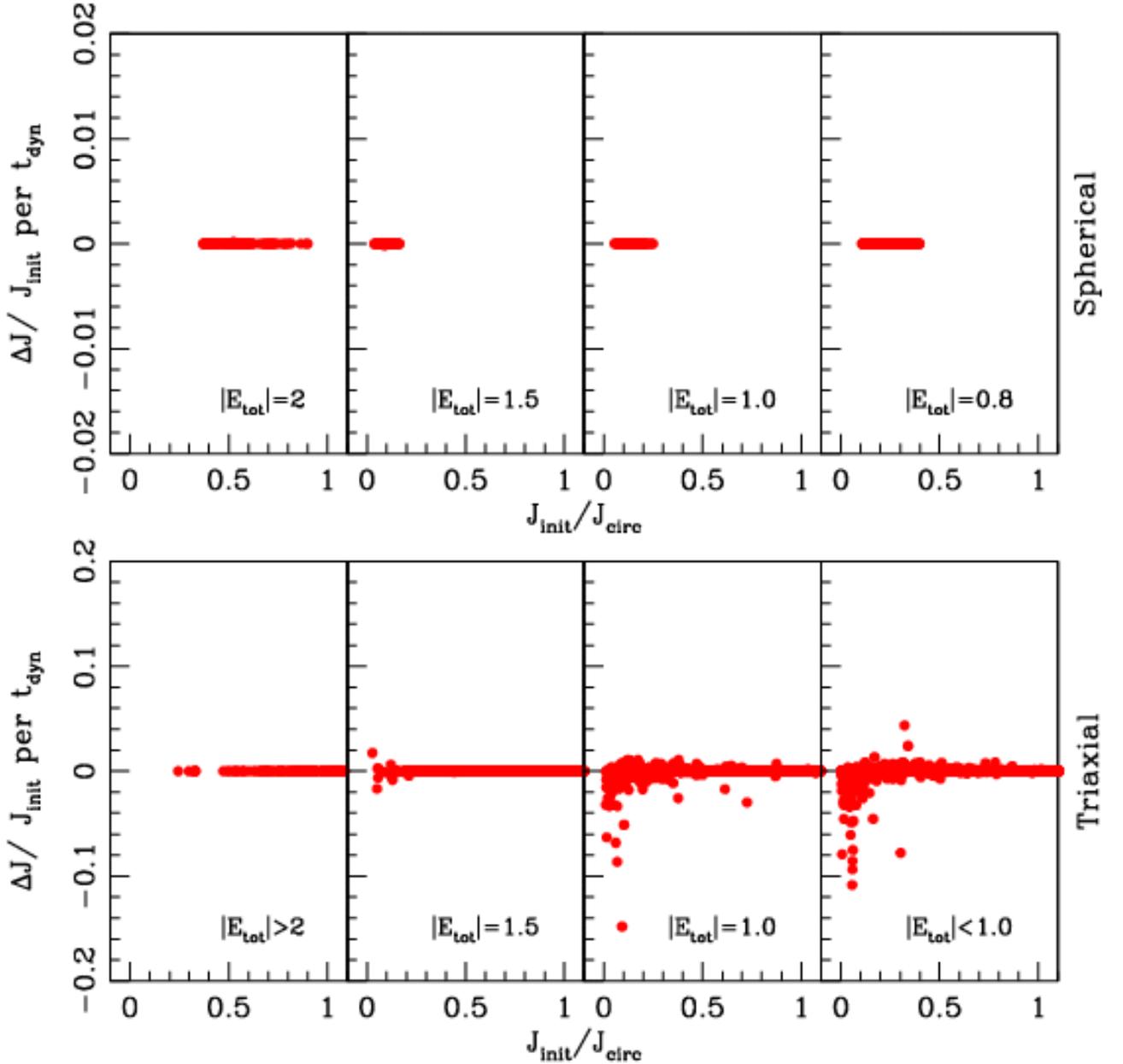,height=500pt, width=500pt}
\caption{The change in the angular momentum per dynamical time as a function of the normalised angular momentum $\kappa$ for several energies. The points 
represent all the orbits in our course-grained sample of the mass density.
In the triaxial potential (bottom), many orbits change
their total angular momenta significantly per dynamical time. Note the 
vertical scale for the spherical model (top) is 10 times smaller  }
\label{fig:angmomall}
\end{figure}

\begin{figure}
\epsfig{file=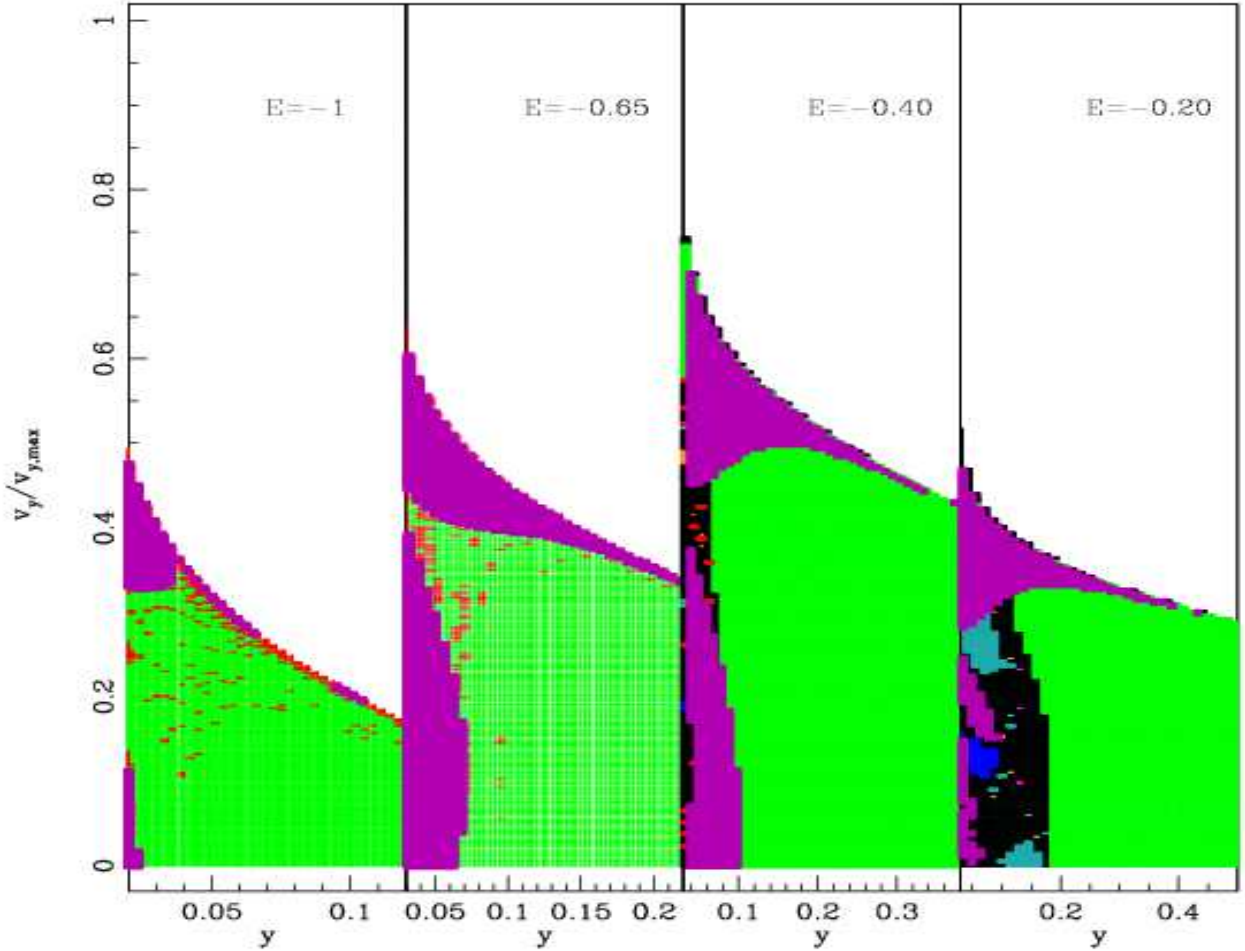,height=400pt, width=500pt}
\caption{The surface of section for several energies. Different colors 
represent different orbit types. Red are chaotic orbits, black are boxes, 
blue are fish, cyan are pretzels, orange are 5:4 resonances. Overlaid in 
purple are the orbits that change angular momentum the most per dynamical 
time of a circular orbit at that energy (the top 10 percent 
in each energy bin are plotted). The orbits that typically move in angular momentum 
the most are boxes and chaotic orbits.}
\label{fig:sos}
\end{figure}

\begin{figure}
\epsfig{file=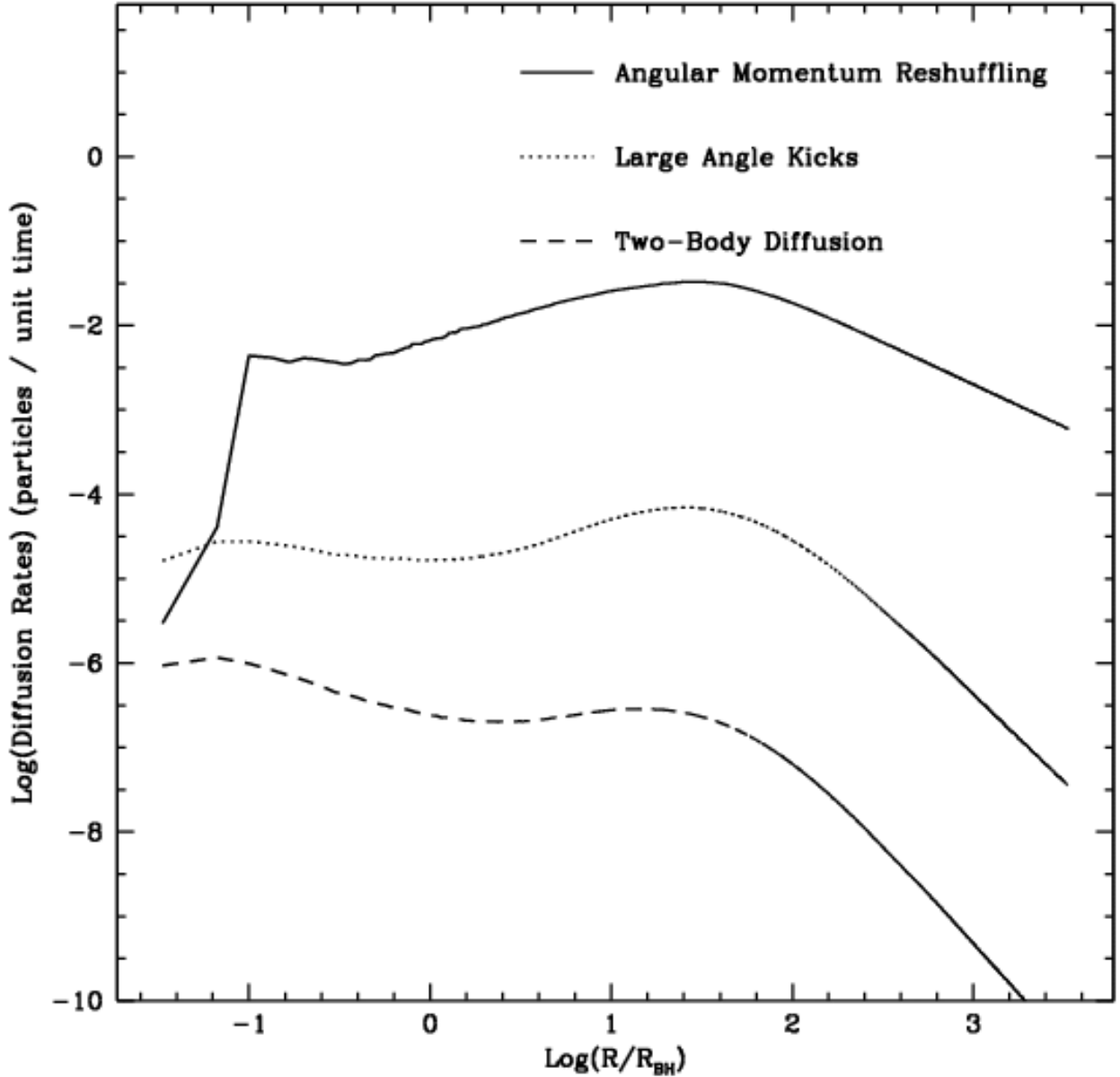, height=500pt, width=500pt}
\caption{The rate of depletion and replenishment of the loss cone as a function 
of radius for all particles in the simulation in model units. Here, the 
particles trace the density coarsely, and each particle represents many 
stars. When scaled to the the Milky Way bulge, the average particle 
in our simulation represents about $800 M_\odot$, and the dynamical time 
at $r_{\rm BH}$ is $\sim 5 \times 10^4$ years. }
\label{fig:loss}
\end{figure} 

\begin{figure}
\epsfig{file=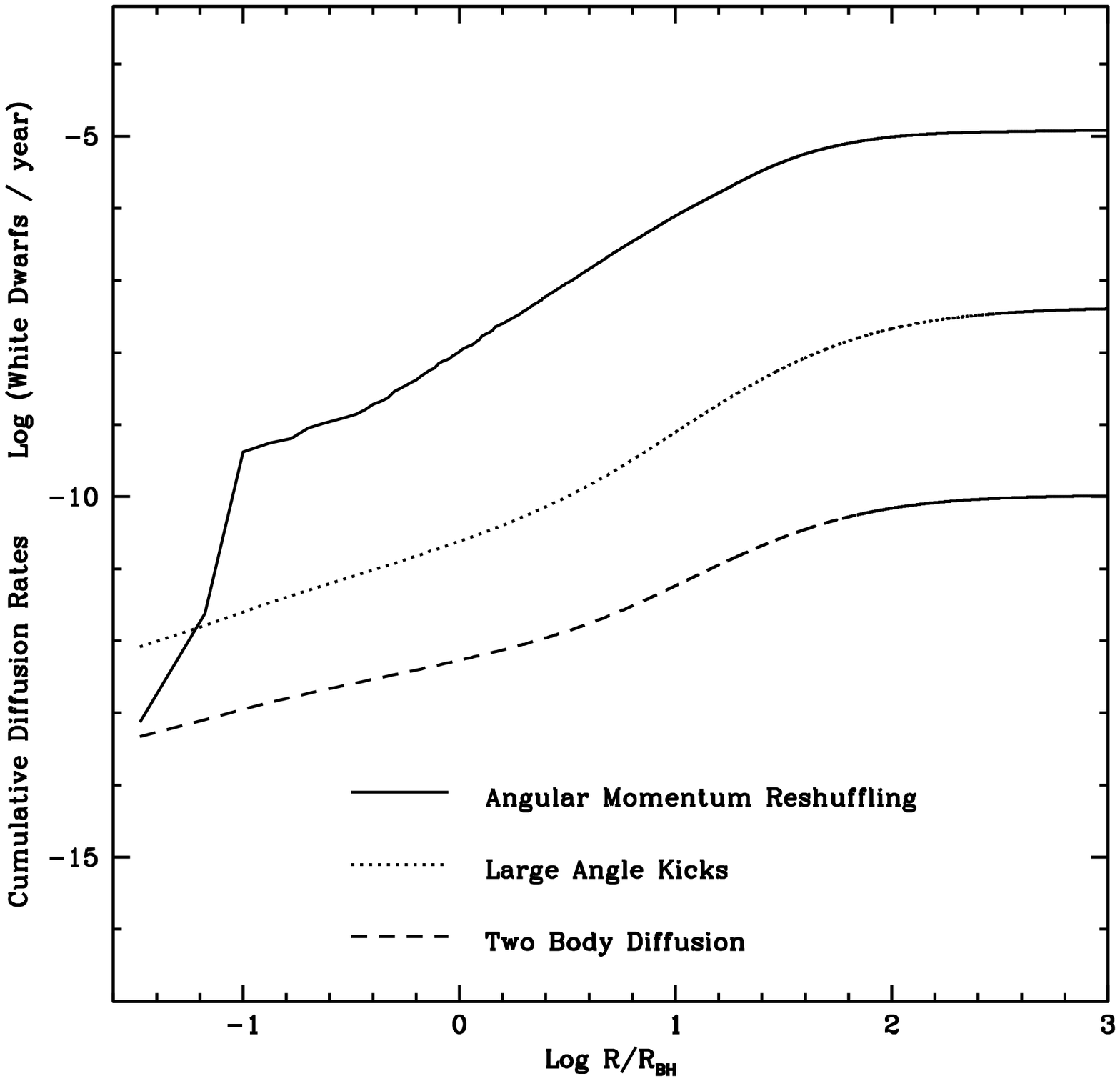, height=500pt, width=500pt}
\caption{The cumulative rate of depletion and replenishment of the 
loss cone as a function of radius for white dwarfs when the model is 
scaled to the physical parameters of the Milky Way bulge. We assume the 
particle 
mass distribution of the simulation maps the smooth mass distribution of the 
galaxy model, and that all stars were formed in a single burst 10 Gyr ago
with a Scalo IMF. White dwarfs are assumed to form from stars with initial 
masses between $1-8 M_\odot$.}
\label{fig:loss2}
\end{figure} 

Figure 4 shows the rate of flow into the 
capture and inspiral loss cone for particles in the simulation under different
processes. Note that the rate particles stream into the loss cone by
this triaxial scrambling process is much larger than the rate in which 
they are lost by 2-body relaxation and large angle scattering combined. And, 
since technically 
every orbit can participate in reshuffling the phase space of the loss cone,
we anticipate that the loss cone reservoir in a triaxial potential 
is always filled. Again, this is because
there is a large amount of flow both in an out of the loss cone
at large radii. Normally, this would be symmetric, but the SMBH
provides a sink that captures an extremely small fraction of the orbits
that just happen to be at the right phase, thereby keeping the loss cone full.

\section{Astrophysical Capture Rate}

In the previous section, we defined a capture and inspiral loss cone
and showed that, in a triaxial potential, this region is 
everywhere full because orbits are capable of streaming though angular 
momentum space.
The capture loss cone contains a smaller subset of particles with
trajectories that slowly spiral into the black hole, which generate clean 
Extreme Mass Ratio Inspiral signals that LISA will detect. 
There is some indication that a significant fraction of stars within the 
capture loss cone may plunge into the SMBH so quickly that LISA will be 
unable to detect an inspiral signature (Hopman \& Alexander 2005). This study,
however, only considered diffusion in the pinhole regime, and it is likely 
that large angle scattering is critical in this part of the loss cone.
Most studies have equated the astrophysical capture rate in the 
Milky Way and M32 to the LISA rate (within a factor of a few). We make 
the same assumption, but caution that the fraction of captures LISA can 
actually detect per galaxy may be small.

We can estimate the rate of capture for different 
compact object populations for a particular galaxy by converting the 
mass density of particles in our simulation into a number density of the 
compact objects of interest. To do this we assume that the mass density
or our coarse-grained model maps the mass density of the galaxy, after
scaling the size and total mass. We assume the mass density 
results from a single burst of star formation 10 Gyr ago, and that
the number of stars per unit mass follows a Scalo IMF (Scalo 1986). We further assume
that white dwarfs formed from all the stars between $ 1-8 M_\odot$ and that
stellar mass black holes formed from stars between $20-300 M_\odot$. This
calculation neglects the effects of mass segregation and multiple episodes of
 star formation, both of which should increase the rate of low mass black hole
captures significantly.
Table 1 presents the white dwarf and stellar
mass black hole capture rate for the Milky Way bulge and for M32, and figure 5
shows the cumulative rate of capture 
for white dwarfs in the Milky Way bar. We find that the 
total large angle scattering and diffusion capture rate for white dwarfs 
is $O(10^{-7})$ per year in the Milky Way bulge, consistent with previous work. The rate of 
captures from angular momentum reshuffling, however, is much 
larger: $O(10^{-5})$ white dwarfs per year. As a rough consistency check,
if a the dynamical time of a typical capture orbit were  $10^4$ years, this 
capture rate suggests capture occurs once every 10 dynamical times at 
that energy, which would yield a reasonable occupation fraction of $10\%$; 
in other words, the flow only captures $10\%$ of the available orbits with 
these parameters.
 
This rate is robust for this 
process only as well as our galaxy model matches the Milky Way inner regious.
The most glaring difference is that a bar is a rotating triaxial 
ellipsoid, which hosts more chaotic orbits, but fewer real boxes. Consequently,
it is
not immediately clear how the rate should be adjusted to account for bulk 
rotation, but the capture rate in the Milky Way bar 
could change by an order of magnitude or more. We plan to study this with 
detailed  N-body simulations in the future.

\begin{table*}
\centering
\caption{Astrophysical Capture Rates and Physical Scaling} 
\begin{tabular}{@{}ccccccc@{}}
\hline
\\
Galaxy 
& $M_\bullet  (M_\odot) $ 
&
& $M_{\rm bulge}  (M_\odot) $
& Effective Radius (pc)
& White Dwarf 
& Black Hole \\
Model &&&&&Captures   (${\rm yr^{-1}}$) & Captures (${\rm yr^{-1}}$)     \\
\\
\hline\hline
\\
M32 & $3\times10^6 $ & &$3\times10^9 $ & 200     & $5\times10^{-6} $ & $9\times10^{-9} $ \\
Milky Way bulge &  $3\times10^6 $ & &$2\times10^{10} $ & 2000     & $1\times10^{-5} $ & $2\times10^{-7} $ \\
\\
\hline

\label{tab:simparams}
\end{tabular} 
\end{table*}

\section{LISA Signals from Intermediate Mass Black Hole Inspirals}

\begin{figure}
\epsfig{file=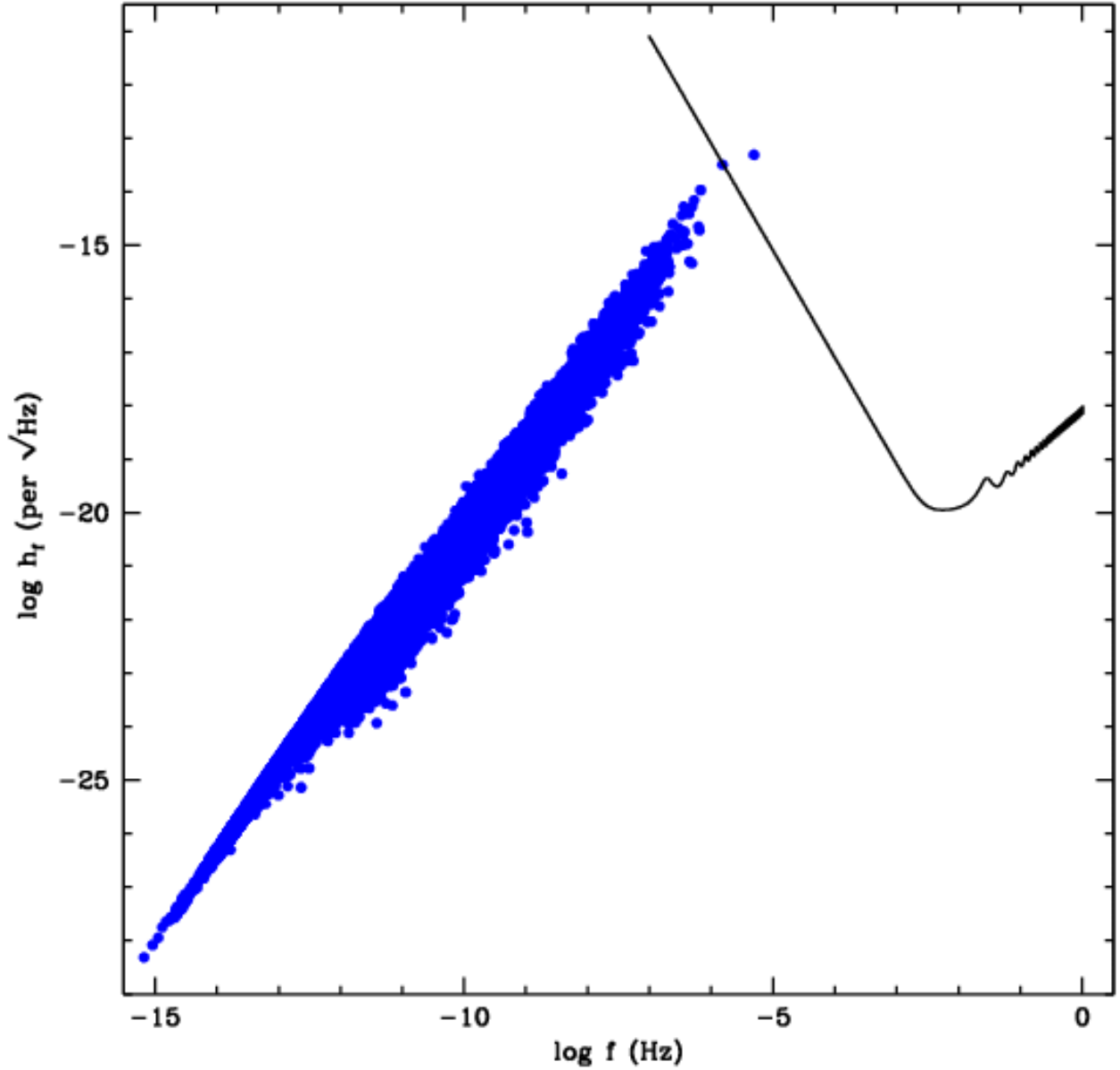, height=500pt, width=500pt}
\caption{The estimated gravitational wave strength as particles pass pericenter in the simulation. The particles were scaled in mass such that the total 
mass is equal to the Milky Way bulge, so that each particle is approximately
$800 M_\odot$, and the strain is calculated assuming the SMBH is 8 kpc
from the detector. Each simulation particle normally represents many stars,
so any IMBH-SMBH rate inferred from this figure would be unphysical. However,
if there were IMBHs in our Milky Way bulge, the centrophilic orbits in the bar would pass close enough to the SMBH to yield LISA burst signals.}
\label{fig:lisafig}
\end{figure} 

For each particle, we determined the gravitational wave signal during 
close encounters with the SMBH. Very near the SMBH, we can approximate 
the orbit as Keplerian; for this equivalent Keplerian orbit, the time 
to travel through the arc described by the latus rectum yields the 
frequency of interaction, while the eccentricity yields the power radiated 
in gravitational waves (Peters \& Mathews 1963):

\begin{equation}{
{P} = {{{8}\over {15}} {{G^4} \over {c^5}} {{{m_1}^2 {M_{\bullet}}^2 (m_1 +M_{\bullet})} \over {a^5 (1-e^2)^5}} (1+e {\rm cos}\theta)^4 \big[12(1+e {\rm cos}\theta)^2 + e^2 {\rm sin}^2 \theta \big]}.}
\end{equation}

The power is related to the strain amplitude in the following way:

\begin{equation}{
{h_f} = { \sqrt{{{4 G} \over {16 c^3 \pi^2 {f_{\rm orb}}^2}} {{P} \over {D^2}}}}.}
\end{equation}

If we scale the model to the Milky Way bulge and allow the particle mass 
to set the mass of the interacting particle ($800 M_\odot$), then our 
simulation produces two highly eccentric Intermediate Mass Black Hole 
(IMBH)-SMBH encounters with a signal to noise ratio of approximately 1000 
(Figure 6). These orbits pass by the SMBH with eccentricities $> 0.99$,
spending only $10^5$ seconds to fly by the SMBH at pericenter distances of
about 200 AU. This could be a relatively unexplored source of
LISA burst signals, assuming that the spacecraft will have the expected 
sensitivity in the $10^{-5}$ Hz band. These encounters are more likely 
with a triaxial 
model, due to 
the increased fraction of centrophilic orbits, though this begs the question 
of how one produces an $O(10^3) M_\odot$ IMBH in the first place 
(Gebhardt et al. 2000b; Gerssen et al. 2003; Taniguchi et al. 2000, Miller \& 
Hamilton 2002). However, even if these were more common 10 $M_\odot$ stellar
mass black holes, they would still produce a LISA burst signal as they 
pass through pericenter.

\section{Implications for Binary Black Hole Mergers}

During a galaxy merger, each 
SMBH sinks to the center of the new galaxy potential due to 
dynamical friction and eventually becomes a bound SMBH binary (SMBBH).
Dynamical friction still shrinks the SMBBH orbit until the binary is hard;
thereafter, further decay is mediated by 3-body scattering with the ambient
stellar background until the SMBBH becomes so close that the orbit can lose
energy via gravitational wave emission, and the SMBBH can presumably coalesce.

The SMBBH loss cone is larger than for single SMBH, since stars must only
pass within the separation a of the SMBBH, where $a = G (M_1+M_2) /8 \sigma_\star^2 \sim 0.05$ parsec 
for two equal $10^6 M_\odot$ SMBHs.
Unfortunately, once the SMBBH has interacted with all the stars
within its loss cone, the orbit decay stalls within the center of
the galaxy with a separation $\sim$ parsec.

This is known as the 'final parsec problem': the orbit of a SMBBH in 
idealized galaxy models never decays enough to allow gravitational wave 
emission to drive the system to coalesce. 
The behavior of stars in angular 
momentum space is deeply connected to this problem, since an ample supply 
of low angular momentum stars is tantamount to providing a fresh reservoir 
of orbits within the loss cone.

Our results for the refilling rate of the capture and inspiral loss cone 
imply that it is always full. In other words, there are many more particles that
can flow through the capture and inspiral loss cone than can be depleted by gravitational
encounters. Hence in principle, the larger loss cone for 
3-body scattering by a binary black hole can also remain full in a triaxial 
potential. We are developing a fully self-consistent n-body model
with $10^7$ particles to explore the shape, time dependence, and
phase space content of the SMBBH loss cone explicitly.

We note that our triaxial galaxy model was constructed from 
an equilibrium spherical potential; if SMBBH are formed as the byproduct
of major galaxy mergers, they likely inhabit post-merger remnant galaxies that
are significantly more flattened ($c/a=0.7$ for our model versus $c/a=0.4-0.6$ 
for the typical equal-mass merger remnant (Statler 2004)). We are determining 
how the loss cone refilling rate depends on the degree of flattening for
a number of equilibrium triaxial models. We are also studying the capture rate
in high resolution galaxy models derived from cosmological n-body simulations
where the galaxy is not necessarily in equilibrium; in this case, global 
potential fluctuations, triaxiality, and the occasional substructure 
merger can all play a role in replenishing the loss cone.

\section{Summary}

At relatively large radii, a significant fraction of orbits in a 
triaxial galaxy stream so rapidly in angular momentum space that they replenish a
loss cone as fast as it is depleted. Hence, we have shown that, in principle,
loss cones in a triaxial galaxy may always be full at all energies.
Nearly every orbit in a triaxial potential participates in this rapid 
angular momentum
flow, reshuffling about $5\%$ of the phase space content near 
$r_{\rm inf}$ over 1000 dynamical times, or on a timescale of 
$10^7$ years for the Milky Way. Well inside $r_{\rm inf}$ the potential is 
rounder, and most orbits that make significant changes 
in angular momentum are chaotic. We found that the loss cone
can be replenished through this process even deep in the potential, where
the figure is nearly spherical.

When we consider angular momentum flow for orbits triaxial potential, we find that the capture rate 
is several orders of magnitude higher than spherical astrophysical capture 
estimates at nearly every radius.
If our astrophysical white dwarf capture rate for 
the Milky Way of $10^{-5}$ per year is correct, it implies that up to
$20 \%$ of our SMBH could have been grown from white dwarf accretion alone.
We emphasize that the fractional number of white dwarfs accreted is
still quite small: though about $25\%$ of the white dwarf population 
passes through pericenter in a orbital time (by definition), only about 
1 in 100000 are captured by the SMBH in our model.

Since dark matter halos are thought to be triaxial in shape, then 
if there is dark matter in galactic central cusps,
this angular momentum flow can also move dark matter onto 
an analogous, though much narrower, capture loss cone. Some estimates
indicate that the density of dark matter dominates the stellar potential
at $0.001$ parsec (e.g. Bertone \& Merritt 2005).
If dark matter is less effective at scattering, particularly if it is 
not self-interacting, triaxiality may be an excellent mechanism to drive 
significant amounts of dark matter into the SMBH.

The capture and inspiral rate for an eternally full loss cone in a 
spherical galaxy has often been used as an upper limit, with the argument 
that a full loss cone cannot be more full. Note, however, that since 
triaxial galaxies have many more centrophilic orbits, the distribution 
function of a full triaxial loss cone can be 'supersaturated' compared to the
full loss cone for a spherical isotropic galaxy. This could occur even the 
orbits were only able to change angular momentum via 2-body processes.

The surprisingly large replenishment rates imply large tidal disruption 
rates as well, since the tidal loss cone width
is about 1000 times larger than the capture and inspiral loss cone. This 
suggests tidal disruption rates as high as $1 \times 10^{-3}$ stars per year 
in a triaxial system with the same $\rho(r)$ as M32.  In fact, this tidal 
disruption rate is consistent with 
what is inferred by the large amplitude X-ray outbursts observed at the 
centers of galaxies (Donley et al. 2002). Our inferred tidal disruption rate
is rather large when our model is scaled to the Milky Way bar; however, it may
be true that the angular momentum reshuffling process is less efficient
in rotating bar potentials. Our numerical models
also suggest that captures and tidal disruption events can originate 
from radii much larger than what is typically assumed; most centrophilic 
orbits in our model had apocenters much larger than 100 pc and though each 
centrophilic orbit contributes very little to the total rate,
there are enough centrophilic orbits in our triaxial model to influence the 
capture or tidal disruption rate at large radii. 

We have shown that triaxiality can increase the capture rate of 
single stars, but it may be even more important for binary systems.
Binaries are funneled from large distances to the SMBH along
centrophilic orbits in a triaxial galaxy. A close 
pericenter pass tidally separates the binary, with one member of the binary 
captured by the SMBH, and one ejected at high velocity. Since tidal separation 
can occur without significant energy loss, the typical
captured star from a binary tidal separation can have an apocenter
hundreds of times larger than a single stellar capture; this may make 
binary interactions a much more efficient capture mechanism 
(Miller et al. 2005), leading to a yet higher EMRI rate.
In addition, the binary EMRI signal itself will be unique. When capture 
occurs at such large separations from
the SMBH, there is ample time for the orbit to circularize by the
time it becomes detectable by LISA. Hence, there may be two classes of EMRI: 
the high eccentricity
inspiral of a single star, and the zero eccentricity inspiral of a tidally
separated binary. Comparing the ratio of the signals can probe the
structure, stellar content, and recent kinematic history
of the central regions of galaxies.

\section*{Acknowledgments}

We would like to acknowledge the support of the Center for 
Gravitational Wave Physics, which is funded by the National 
Science Foundation under the cooperative agreement PHY 01-14375. 
This work was completed with the support of a grant from the NSF, PHY 02-03046,
 and,from NASA, ATP NNG04GU99G.
This manuscript was written at the Aspen Center for Physics 
and KHB would also like to thank the Center and the other participants 
for their hospitality.


{}

\label{lastpage}

\end{document}